\documentstyle[preprint,aps]{revtex}

\begin{document}
\draft
\date{\today}
\preprint{{\Large ${\rm WU-B-94-16} \atop {\rm RUB-TPII-04/94}$}}
\title{NEUTRON FORM FACTOR: SUDAKOV SUPPRESSION AND INTRINSIC
       TRANSVERSE SIZE EFFECT
      }
\author{J. BOLZ, R. JAKOB and
        P. KROLL\footnote{E-mail address:
                          kroll@wpts0.physik.uni-wuppertal.de}
       }
\address{Fachbereich Physik          \\
         Universit\"at Wuppertal     \\
         D-42097 Wuppertal, Germany
        }
\author{M. BERGMANN and
        N.~G. STEFANIS\footnote{E-mail address:
                                nicos@hadron.tp2.ruhr-uni-bochum.de}
       }
\address{Institut f\"ur Theoretische Physik II  \\
         Ruhr-Universit\"at Bochum              \\
         D-44780 Bochum, Germany                \\
         }
\maketitle
\newpage
\begin{abstract}
Two recently proposed concepts to improve the perturbative calculation
of exclusive amplitudes, gluonic radiative corrections (Sudakov factor)
and confinement size effects (intrinsic transverse momentum) are
combined to study the neutron magnetic form factor in the space-like
region. We find that nucleon distribution amplitudes modelled on the
basis of current QCD sum rules indicate overlap with the existing
data at the highest measured values of momentum transfer.
However, sizeable higher-order perturbative corrections (K-factor)
and/or higher-twist contributions cannot be excluded, although
they may be weaker than in the proton case.
\end{abstract}
\narrowtext
\newpage 

In a recent paper~\cite{BJKBS94} we have studied the space-like proton
form factor within a theoretical scheme proposed by Li and
Sterman~\cite{LS92}, which takes into account gluonic radiative
corrections in the form of a Sudakov factor~\cite{BS89}. This scheme
naturally generalizes the standard hard scattering picture (HSP) of
Brodsky and Lepage~\cite{LB80}---commonly used to calculate exclusive
reactions within perturbative QCD---by taking into account the
transverse momentum of the partons.

A major point in our proton form-factor analysis was to show that
proper treatment of the
$\alpha _{s}$-singularities demands the imposition of an appropriate
infrared (IR) cut-off to render the form-factor calculation both finite
and insensitive to the inclusion of the soft region of phase space.
This is in contrast to the pion case~\cite{LS92}, where a ``natural''
IR cut-off appears in the form of the interquark separation.
Considering in detail optional IR cut-off
prescriptions~\cite{Li93,Hye93,SS94}, we found that maximum IR
protection is provided by introducing as a common IR cut-off in the
Sudakov (suppression) factor the maximum interquark separation
(``MAX'' prescription~\cite{BJKBS94}).
The underlying physical idea is the following:
One expects that because of the color neutrality of a hadron, its quark
distribution cannot be resolved by gluons with a wavelength much larger
than a characteristic interquark separation scale $\tilde{b}_{l}$.
Thus, gluons with wavelengths large compared to the (transverse) hadron
size probe the hadron as a whole, i.e., in a color-singlet state and
decouple. As a result, quarks in such configurations act coherently and
therefore (soft) gluon radiation is dynamically inhibited.

The ``MAX'' presription not only suffices to suppress the
$\alpha _{s}$-singularities, but also preserves the finiteness of the
integrand of the expression for the form factor, even when
renormalization-group (RG) evolution of the wave function is included.
An additional bonus of the ``MAX'' prescription is that the proton form
factor saturates, i.e., becomes insensitive to the contributions from
large transverse separations. Admittedly, little confidence is put in
perturbative treatments of the large-distance region, so that saturation
of the form factor at transverse distances as low as possible is a
prerequisite for a self-consistent perturbative calculation.

In \cite{BJKBS94} we have pointed out that the recent numerical analysis
by Li~\cite{Li93} of the proton form factor has serious drawbacks:
(i) The cancellation of the $\alpha _{s}$-singularities in the region
$\tilde{b}_{l}\Lambda _{\text{QCD}} \simeq 1$ ($x_{l}$ fixed) is
incomplete for different $\tilde{b}_{l}$, amounting to uncompensated
singularities of the form
\begin{equation}
  \sim \ln \left(
                 \frac{1}{\tilde{b}_{l}\Lambda _{\text{QCD}}}
           \right)^{\kappa}.
\label{eq:sing}
\end{equation}
(ii) There is no saturation, meaning that the main form-factor
contributions are accumulated in the ``forbidden'' soft region.

A second element of our approach in \cite{BJKBS94} was the incorporation
of the intrinsic transverse momentum in the proton wave function,
following a previous work by two of us~\cite{JK93} on the pion form
factor. The intrinsic transverse momentum reflects confinement-size
effects~\cite{Kro93} and improves the saturation behavior of the form
factor. As a consequence of these effects (Sudakov factor and intrinsic
transverse momentum), the self-consistent perturbative contribution to
the proton form factor turns out to be reduced by at least a factor of
two compared to the existing experimental data
(see, e.g.,~\cite{Arn86,Sil93}).
This is true for a variety of nucleon distribution amplitudes (DA),
recently determined by two of us~\cite{BS93,BS94} on the theoretical
basis of QCD sum rules~\cite{COZ89,KS87}.

In the present work we extend this type of analysis to the neutron
magnetic form factor. While in our previous work the focus was on the
self-consistent implementation of the Sudakov factor and the proper
identification and matching of the scales involved, the subject of the
current effort will be on the phenomenological side. In particular, we
consider observables involving the ratio of the neutron to the proton
form factor $G_{M}^{n}/G_{M}^{p}$. It has been discussed
in \cite{Ste89} (and previous references cited therein) and more recently
in \cite{Ste93a} that proposed model distribution amplitudes for the
nucleon can be classified according to this ratio (an observable
quantity) and the theoretical parameters $B_{4}$ (projection
coefficient on the corresponding eigenfunction of the nucleon evolution
equation) and the ``hybridity'' angle $\vartheta$, as detailed
in \cite{BS94}.
One of the crucial questions posed in the present work is whether the
emerging pattern of solutions to the sum rules, found within the standard
HSP~\cite{LB80,COZ89} to constitute a smooth and finite ``orbit'' in
the $(B_{4},-G_{M}^{n}/G_{M}^{p})$ plane, pertains to the inclusion of
transverse-momentum contributions.

The starting point of our analysis is to consider the neutron magnetic
form factor within the modified HSP:
\begin{equation}
  G_{M}^{n}(Q^{2})
=
  \frac{16}{3}
  \int_{0}^{1}[dx][dx']
  \int_{}^{}\frac{d{}^{2}b_{1}}{(4\pi )^{2}}
            \frac{d{}^{2}b_{2}}{(4\pi )^{2}}
  \sum_{j=1}^{2}\, \hat{T}_{j}(x,x',\vec{b},Q,\mu )
  \hat{Y}_{j}^{n}(x,x',\vec{b},\mu _{F}) \,
  {\rm e}^{-S_{j}},
\label{eq:G_M^n(b)}
\end{equation}
with
$[dx]=dx_{1}dx_{2}dx_{3}\delta (1-\sum_{}^{}x_{i})$; $x_{i}$ being the
momentum fractions carried by the valence quarks.
The Fourier-transformed hard scattering amplitudes are given by
\begin{equation}
  \hat{T}_{1}
=
  \frac{8}{3}\,C_{\text{F}}\,\alpha _{s}(t_{11}) \alpha _{s}(t_{12})
  K_{0}
       \left(
             \sqrt{(1-x_{1})(1-x_{1}')}\,Qb_{1}
       \right)
  K_{0}\left(
             \sqrt{x_{2}x_{2}'}\,Qb_{2}
       \right),
\label{eq:FourierT_1}
\end{equation}
\begin{equation}
  \hat{T}_{2}
=
  \frac{8}{3}\,C_{\text{F}}\,\alpha _{s}(t_{21}) \alpha _{s}(t_{22})
  K_{0}
       \left(
             \sqrt{x_{1}x_{1}'}\,Qb_{1}
       \right)
  K_{0}
       \left(
             \sqrt{x_{2}x_{2}'}\,Qb_{2}
       \right),
\label{eq:FourierT_2}
\end{equation}
where the $K_{0}$ are modified Bessel functions of order $0$ and $b_{l}$
denotes the length of the corresponding transverse-distance vector.

The renormalization scale is chosen in such a way that each hard
gluon refers to its own individual momentum scale $t_{ji}$ to be used in
the argument of the corresponding $\alpha _{s}$.
The $t_{ji}$ is defined as the maximum scale of either the longitudinal
momentum or the inverse transverse separation, associated with each of
the gluons: viz.
\begin{eqnarray}
& t_{11} &
=
  {\rm max} \left[
                  \sqrt{(1-x_{1})(1-x_{1}^{\prime})}\,Q, 1/b_{1}
            \right],
\nonumber \\
& t_{21} &
=
  {\rm max} \left[
                  \sqrt{x_{1}x_{1}^{\prime}}\,Q, 1/b_{1}
            \right],
\nonumber \\
& t_{12} &
=
  t_{22}
=
  {\rm max} \left[
                  \sqrt{x_{2}x_{2}^{\prime}}\,Q, 1/b_{2}
            \right].
\label{eq:t_ij}
\end{eqnarray}

Since the hard scattering amplitudes depend only on the differences of
initial and final state transverse momenta, there are only two
transverse separation vectors, namely those between quarks 1 and 3 and
between quarks 2 and 3:
$\vec{b}_{1}\;(=\vec{b}_{1}')$, $\vec{b}_{2}\;(=\vec{b}_{2}')$.
Accordingly, the transverse separation between quark 1 and quark 2 is
$
  \vec{b}_{3}
=
  \vec{b}_{2} - \vec{b}_{1}.
$

The soft part of the form factor is given by the following expressions
which contain linear combinations of products of the initial and final
state wave functions in the transverse configuration space, weighted by
$x_{i}$-dependent factors arising from the fermion propagators:
\begin{eqnarray}
  \hat{Y}_{1}^{n}
=
  \frac{1}{(1-x_{1})(1-x^{\prime}_{1})}
\Bigl\{  & \phantom{} &
\!\!\!\!\!\!
           -2\hat{\Psi} ^{\star\prime}_{123}\hat{\Psi} _{123}
           -2\hat{\Psi} ^{\star\prime}_{132}\hat{\Psi} _{132}
          +  \hat{\Psi} ^{\star\prime}_{231}\hat{\Psi} _{231}
          +  \hat{\Psi} ^{\star\prime}_{321}\hat{\Psi} _{321}
\nonumber \\
& - &        \hat{\Psi} ^{\star\prime}_{231}\hat{\Psi} _{132}
            -\hat{\Psi} ^{\star\prime}_{132}\hat{\Psi} _{231}
            -\hat{\Psi} ^{\star\prime}_{321}\hat{\Psi} _{123}
            -\hat{\Psi} ^{\star\prime}_{123}\hat{\Psi} _{321}
  \Bigr\}
\label{eq:Y_1^n}
\end{eqnarray}
\begin{eqnarray}
  \hat{Y}_{2}^{n}
= & \phantom{} &
\!\!\!\!\!\!
  \frac{1}{(1-x_{2})(1-x^{\prime}_{1})}
  \left\{
           \hat{\Psi} ^{\star\prime}_{231}\hat{\Psi} _{231}
         + \hat{\Psi} ^{\star\prime}_{231}\hat{\Psi} _{132}
         + \hat{\Psi} ^{\star\prime}_{132}\hat{\Psi} _{231}
  \right\}
\nonumber \\
& + & \frac{1}{(1-x_{3})(1-x^{\prime}_{1})}
  \left\{
           2\hat{\Psi} ^{\star\prime}_{321}\hat{\Psi} _{321}
         -  \hat{\Psi} ^{\star\prime}_{123}\hat{\Psi} _{123}
         +  \hat{\Psi} ^{\star\prime}_{321}\hat{\Psi} _{123}
         +  \hat{\Psi} ^{\star\prime}_{123}\hat{\Psi} _{321}
  \right\}.
\label{eq:Y_2^n}
\end{eqnarray}
The Fourier transform of the wave function reads
\begin{equation}
  \hat{\Psi}_{123}(x,\vec{b},\mu _{F})
=
  \frac{1}{8\sqrt{N_{\text{c}}!}}
  f_{\text{N}}(\mu _{F})
  \Phi _{123}(x,\mu _{F})
  \hat{\Omega}_{123}(x,\vec{b}),
\label{eq:FourierPsi}
\end{equation}
wherein its intrinsic $k_{\perp}$-dependence is parametrized according to
the Gaussian
\begin{equation}
  \hat{\Omega}_{123}(x,\vec{b})
=
  (4\pi )^{2}
  {\rm exp}
           \left\{
                  - \frac{1}{4a^{2}}
                  \Bigl[
                          x_{1}x_{3}b_{1}^{2} + x_{2}x_{3}b_{2}^{2}
                        + x_{1}x_{2}b_{3}^{2}
                  \Bigr]
           \right\}.
\label{eq:FourierOmega}
\end{equation}
We have used the convenient short-hand notation
$
 \hat{\Psi} _{123}(x,\vec{b})
=
 \hat{\Psi} (x_{1},\vec{b};
             x_{2},\vec{b};
             x_{3},\vec{b})
$
denoting in
$
  \Phi _{123}(x,\mu _{F})
$
the factorization scale of short-and large-distance
contributions by $\mu _{F}$.

The exponentials ${\rm e}^{-S_{j}}$ in (\ref{eq:G_M^n(b)}) are the
Sudakov factors responsible for the effects of gluonic radiative
corrections. They have been calculated by Botts and Sterman~\cite{BS89}
using resummation techniques in the context of the renormalization
group (RG) and having recourse to previous extensive work by Collins,
Soper, and Sterman~\cite{CS81}. The explicit expressions for the
Sudakov exponents are given in \cite{LS92,Li93}.

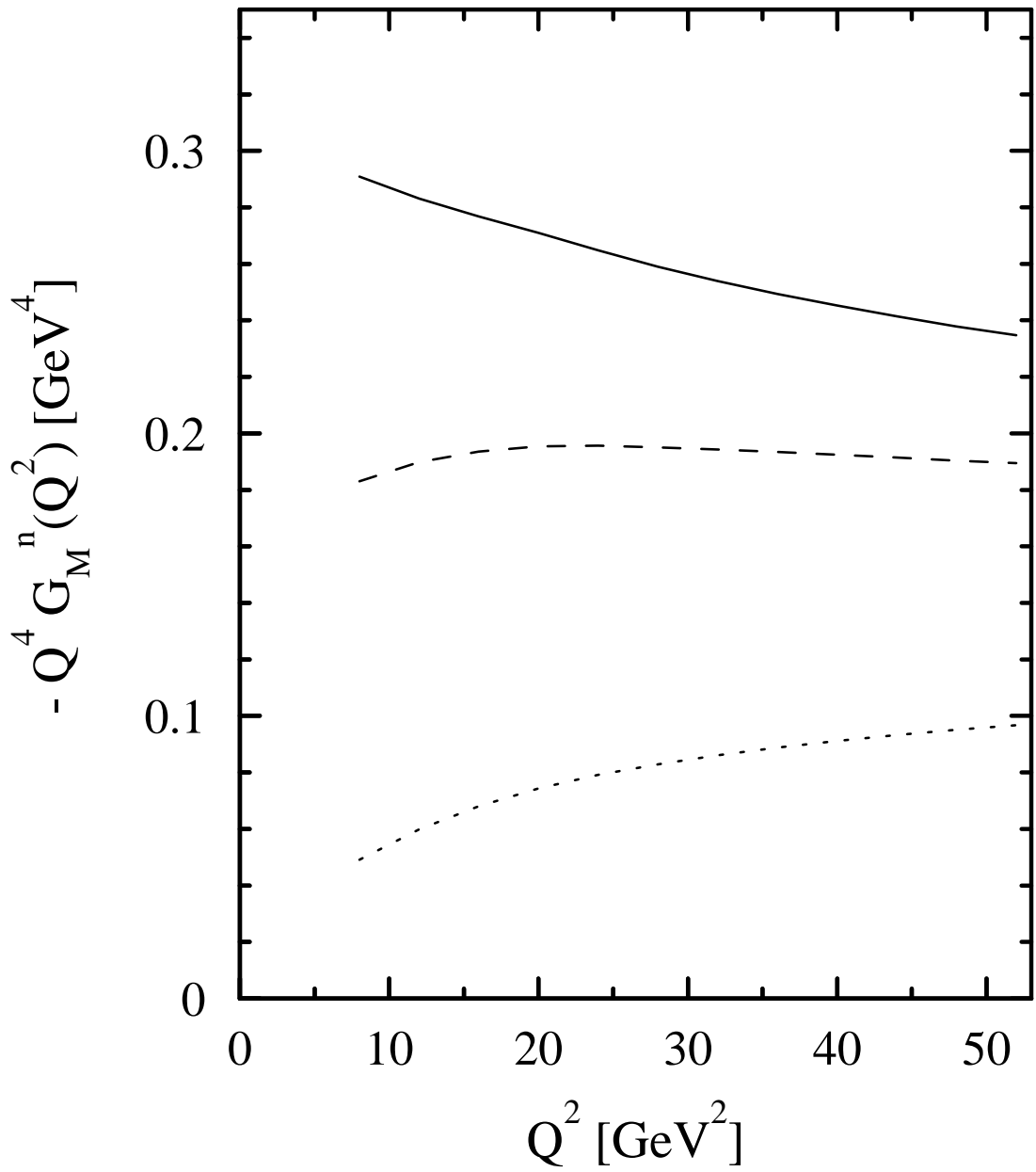
\begin{figure}
\caption[fig:G_M^n(Q^2)]{The influence of the intrinsic transverse
         momentum on the neutron magnetic form factor within the
         modified HSP. The curves shown are obtained for the
         COZ~\cite{COZ89} DA by imposing the ``MAX'' prescription
         including evolution. The solid line represents the results
         without $k_{\perp}$-dependence, whereas the dashed and dotted
         lines are obtained with
         $\langle k^{2}_{\perp}\rangle ^{1/2}=271$~MeV
         and $600$~MeV, respectively.}
\label{fig:G_M^n(Q^2)}
\end{figure}

The analytical and numerical evaluation of the neutron form factor is
performed under the imposition of the ``MAX''
IR-prescription~\cite{BJKBS94} on the Sudakov factor, i.e., setting
\begin{equation}
  \tilde{b}\equiv {\rm max}\{b_{1},b_{2},b_{3}\}
=
  \tilde{b}_{1}=\tilde{b}_{2}=\tilde{b}_{3}.
\label{eq:MAX}
\end{equation}
The results of this calculation are typified by the curves shown in
Fig.~\ref{fig:G_M^n(Q^2)} for the COZ DA~\cite{COZ89} (solid line),
including also the intrinsic transverse momentum in two different ways:
(i) by normalizing the probability $P_{3q}$ for finding three valence
quarks in the neutron to unity (dashed line), which results to
$\langle k^{2}_{\perp}\rangle ^{1/2}=271$~MeV;
and (ii) by setting the value of the r.m.s. transverse momentum equal to
$600$~MeV (dotted line), which implies $P_{3q}=0.042$. Here and below we
throughout use the values $\Lambda _{\text{QCD}}=180$ MeV and
$|f_{N}|=(5.0\pm 0.3)\times 10^{-3}$ GeV${}^{2}$~\cite{COZ89}, the
latter being the value of the nucleon DA at the origin.
The momentum evolution of the DA is RG-controlled---provided the model
DA is satisfying the nucleon evolution equation~\cite{LB80}, which is
true for all DAs we consider in this work. Then the nucleon DA can be
expanded in terms of the eigenfunctions of the one-gluon exchange kernel
to read
\begin{equation}
  \Phi _{123}(x,\mu )
=
  \Phi _{123}^{\text{as}}(x)
  \sum_{n}^{}B_{n}
  \left(\frac{\alpha _{s}(\mu )}{\alpha _{s}(\mu _{0})}
  \right)^{\tilde{\gamma} _{n}/\beta_{0}}
 \tilde{\Phi}_{123}^{n}(x),
\label{eq:Phi}
\end{equation}
where the notations of \cite{Ste89} are adopted and
$
  \Phi _{123}^{as}(x) = 120 x_{1}x_{2}x_{3}
$
is the asymptotic DA. The exponents $\tilde{\gamma}_{n}$ are related to
the anomalous dimensions of trilinear quark operators with isospin $1/2$
(see~\cite{Pes79}) and resemble the $b_{n}$ in the Brodsky-Lepage
notation~\cite{LB80}. Because they are positive fractional numbers
increasing with n, higher-order terms in (\ref{eq:Phi}) are gradually
suppressed. The constants $\tilde{\gamma}_{n}$ are given
in \cite{BJKBS94,Li93}; $\beta_{0}=11-2n_{f}/3=9$ for three flavors.
Under these conditions we may use QCD sum-rule results on the moments
\begin{equation}
  \Phi ^{(n_{1}n_{2}n_{3})}(\mu _{0})
=
  \int_{0}^{1}[dx]x_{1}^{n_{1}}x_{2}^{n_{2}}x_{3}^{n_{3}}
  \Phi _{123}(x,\mu _{0})
\end{equation}
\label{eq:moments}
to constrain the first few expansion coefficients $B_{n}$ (for more
details see \cite{BS94,Ste89,Ste93a}).

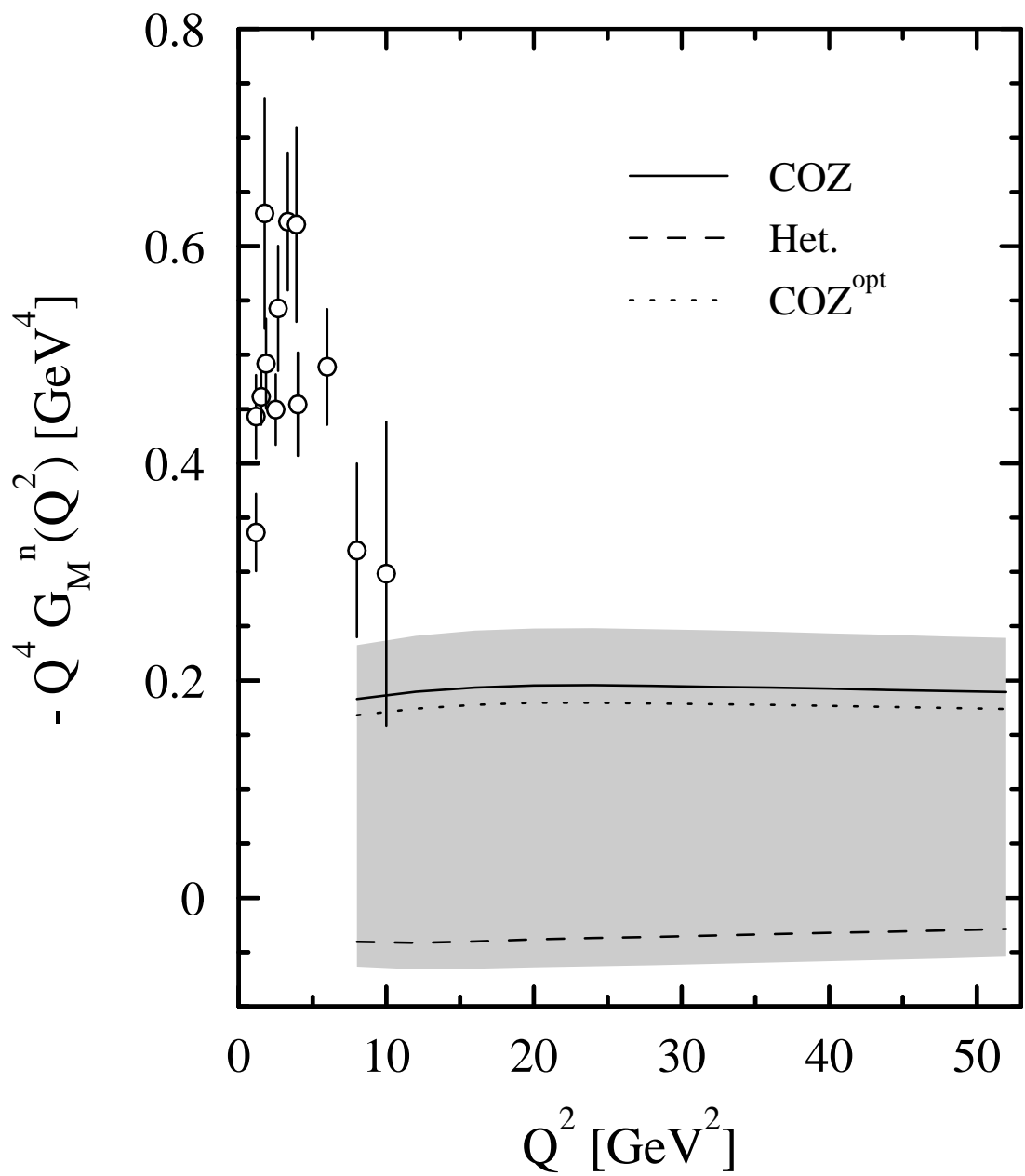
\begin{figure}
\caption[fig:stripn]{The neutron magnetic form factor vs.~$Q^{2}$.
         The theoretical results are obtained using the ``MAX''
         prescription including evolution and normalizing the wave
         functions to unity. The shadowed strip indicates the range of
         predictions derived from the set of DAs determined in
         \cite{BS93} in the context of QCD sum rules (see text).
         The solid (dashed, dotted) line corresponds to the COZ
         (heterotic, optimized COZ) DA (cf.~Fig.~4). The data are taken
         from \cite{Bos92}.}
\label{fig:stripn}
\end{figure}

As can be seen from Fig.~\ref{fig:G_M^n(Q^2)} and comparison with
\cite{Ste93a}, the magnitude of the neutron magnetic form factor with
Sudakov correction is reduced by more than a factor of 2.5 with respect
to the standard HSP.
This reduction is enhanced when the intrinsic transverse momentum is
included and is pertinent to all nucleon DAs~\cite{BS93,BS94} modelled
on the basis of existing QCD sum-rules~\cite{COZ89,KS87} (shaded band in
Fig.~\ref{fig:stripn}).
The upper region of the band is characterized by
COZ-like~\cite{COZ89} DAs, whereas its lower part is associated with the
recently proposed ``heterotic'' DA~\cite{SB93}.
We note that the perturbative contribution becomes self-consistent for
momentum transfers larger than 8 GeV${}^{2}$
(using $\langle k_{\perp}^{2}\rangle ^{1/2}=271$ MeV)
in the sense that at least 50~\% of the result is accumulated in regions
where $\alpha _{s}^2$ is smaller than $0.5$.

It is important to emphasize that the neutron magnetic form factor is
the first process calculated within the modified HSP that yields
predictions which indicate overlap with the existing data~\cite{Bos92},
as can be seen from Fig.~\ref{fig:stripn}.
This tentative agreement occurs at data points corresponding to the
largest momentum transfers measured, where, incidentally, our
theoretical calculations become self-consistent. Therefore, measurements
of the neutron magnetic form factor beyond 10 GeV${}^{2}$ are extremely
important in order to check the validity of the theoretical predictions
in a more quantitative way.

One place to test these results is in the data for the differential
cross sections for elastic electron-proton and electron-neutron
scattering $\sigma_{\text{p}}$ and $\sigma_{\text{n}}$, respectively.
For small scattering angles, where the terms
$\propto \tan^{2}(\theta/2)$
can be neglected, and for large $Q^{2}$, the ratio
$
 \sigma_{\text{n}}/ \sigma_{\text{p}}
$
becomes in a slightly model-dependent way proportional to the
square of the ratio of the neutron to the proton magnetic form
factor.

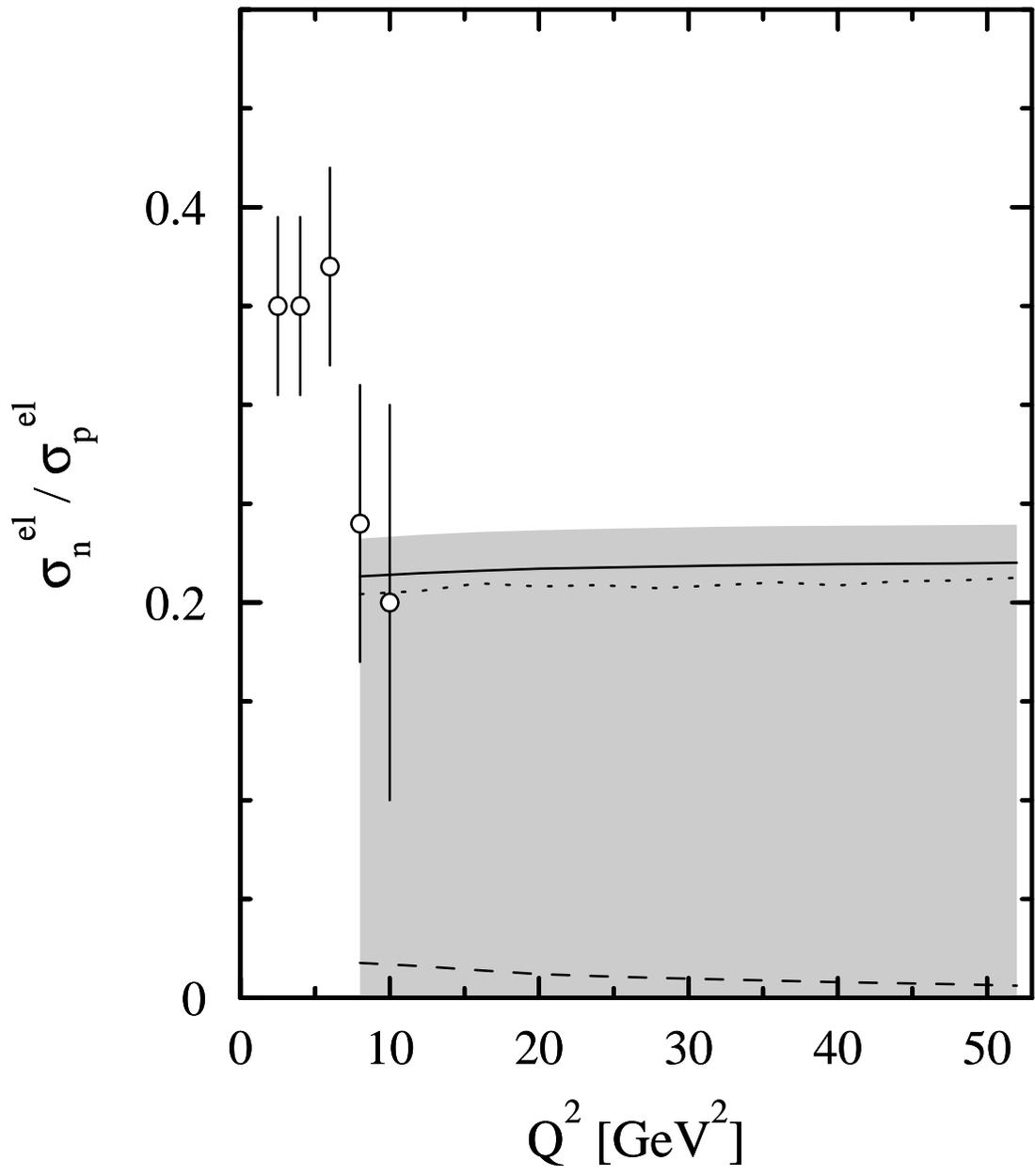
\begin{figure}
\caption[fig:stripsigma]{The ratio
         $\sigma _{n}^{el} / \sigma _{p}^{el}$
         of the differential elastic electron-neutron to
         electron-proton cross section vs.~$Q^{2}$ at scattering angles
         of $10^{\circ}$.
         The shaded area and model DAs for the nucleon correspond to
         those shown in Fig.~2. The data are taken from \cite{Roc92}.}
\label{fig:stripsigma}
\end{figure}

Combining our calculations for the proton~\cite{BJKBS94} with those
presented here for the neutron, we can extract theoretical predictions
for
$
 \sigma_{\text{n}}/ \sigma_{\text{p}}
$
by inputting the same set of model DAs for the nucleon~\cite{BS93,BS94}
as before.
The results are shown in Fig.~\ref{fig:stripsigma} (shaded area) in
comparison with available data~\cite{Roc92}.
{}From this figure we see that the measured values of
$\sigma_{\text{n}}/\sigma_{\text{p}}$ enter the estimated range
already at $Q^{2}\approx 8$~GeV${}^{2}$.
[The corresponding values of the ratio $-G_{M}^{n}/G_{M}^{p}$,
allowed by our analysis, range between -0.2 and 0.5.]
The fair agreement between theoretical predictions and data is partly
deceptive owing to the fact that the self-consistent calculation of the
leading-order perturbative contribution to the proton magnetic form
factor within the modified HSP yields a rather small
value~\cite{BJKBS94}. This missing part of the proton form factor can
arise from many sources, e.g, from a large K-factor and/or higher-twist
contributions and would certainly affect the width of the strip.
Such contributions are also conceivable for the neutron form factor.

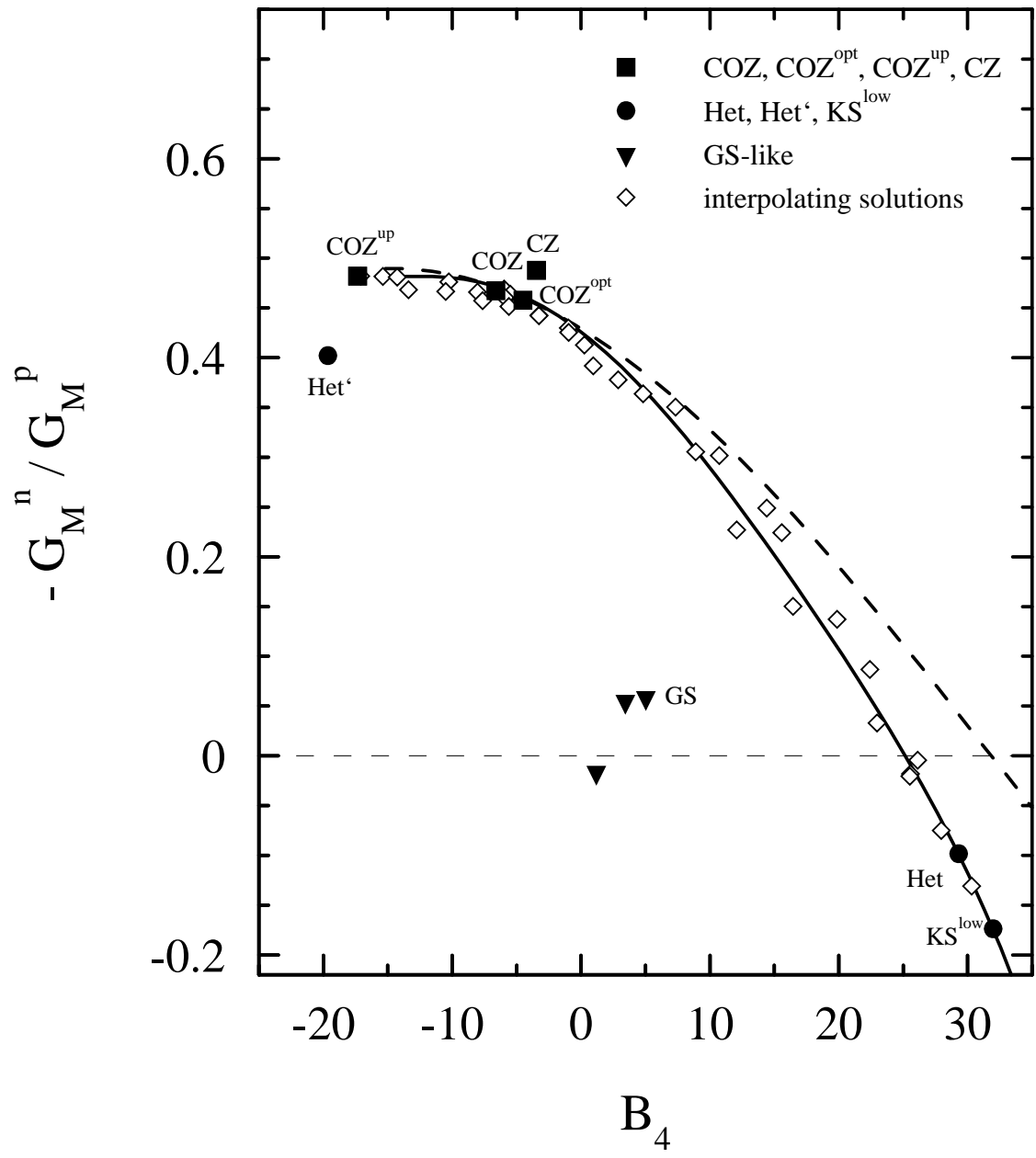
\begin{figure}
\caption[fig:orbit]{Relation between the ratio $R$ of the
         magnetic nucleon form factors and the expansion coefficient
         $B_{4}$ of the Appell polynomial decomposition of the
         nucleon DA, including the effect of the Sudakov factor.
         The results are obtained at $Q^2=30$ GeV$^2$,
         employing the ``MAX'' prescription.
         The superimposed solid line is an empirical polynomial fit
         similar to the original one given in \cite{BS93}. The dashed
         line serves to illustrate the dependence on the momentum scale
         ($Q^{2}=10^{3}$ GeV${}^{2}$).}
\label{fig:orbit}
\end{figure}

In any case it is remarkable that the collective pattern of solutions
to the QCD sum rules \cite{COZ89,KS87}, found within the standard HSP
\cite{BS93,BS94}, pertains to the inclusion of transverse-momentum
contributions comprising the Sudakov factor and those due to the
intrinsic transverse momentum (see Fig. \ref{fig:orbit}). Indeed, the
solutions arrange themselves across an ``orbit'' in the
($B_{4},-G_{M}^{n}/G_{M}^{p}$) plane which is somewhat shifted compared
to the original one. In contrast to the standard HSP version, within the
present context, the ``orbit'' is slightly $Q^{2}$-dependent, as shown
in Fig. \ref{fig:orbit}. The new ``orbit'' at $Q^{2}=30$~GeV${}^{2}$
can be characterized by the empirical relation
$
 -G_{M}^{n}/G_{M}^{p}
=
  0.426 - 9.91 \times 10^{-3} B_{4} - 4.27 \times 10^{-4} B_{4}^{2}
  + 4.59 \times 10^{-6} B_{4}^{3} ,
$
which complies with that found in \cite{BS93}.
The dashed line in Fig.~\ref{fig:orbit} represents a similar fit for
$Q^{2}=10^{3}$~GeV${}^{2}$. We observe that with increasing momentum
transfer, the ``orbit'' within the modified HSP transmutes into that
of the standard HSP. We note that the coefficient $B_{4}$ projects onto
the eigenfunction
$\tilde{\Phi}_{4}(x_{i})$ and hence provides an effective measure to
account for the antisymmetric content of the nucleon DA, since the other
antisymmetric eigenfunctions are offset by this term~\cite{BS93}.

In summary, in this letter the modified HSP has been applied to the
neutron magnetic form factor for the first time. In contrast to other
cases (e.g., pion and proton electromagnetic form factors), the band of
predictions obtained with the set of model DAs for the nucleon indicates
overlap with the experimental data at the largest measured values of
momentum transfer, where the theoretical predictions become
self-consistent.
Nevertheless, it is likely that in order to improve agreement with the
data, several additional contributions have to be included: Perturbative
higher-order corrections may give rise to a rather large K-factor of the
order of 2 multiplying the leading-order contribution, as found for other
large-momentum transfer processes~\cite{Ant83}. However, for the case of
the pion form factor, already existing calculations~\cite{Fie81,DR81} of
the K-factor to one-loop order indicate that with an appropriate choice
of the renormalization point, the actual value of the K-factor is rather
small, i.e., of order unity. On the other hand, still unestimated
contributions due to higher twists are presumably sizeable in the
experimentally accessible region and may also be important.

\acknowledgements
This work was supported in part by the Deutsche Forschungsgemeinschaft
and the Bundesministerium f\"ur Forschung und Technologie FRG under
contract 06WU737.

\newpage 

\newpage 

\end{document}